\documentclass{elsart}
\usepackage{graphicx}
\usepackage{amssymb}
\begin{document}
\begin{frontmatter}
\title{The Velocity of Money in a Life-Cycle Model}
\author{Yougui Wang\corauthref{cor}},
\author{Hanqing Qiu}
\corauth[cor]{Corresponding author. Email: ygwang@bnu.edu.cn(Y.
Wang)}
\address{Department of Systems Science, School of Management, Beijing Normal University, Beijing,
100875, People's Republic of China}

\begin{abstract}
The determinants of the velocity of money have been examined based
on life-cycle hypothesis. The velocity of money can be expressed
by reciprocal of the average value of holding time which is
defined as interval between participating exchanges for one unit
of money. This expression indicates that the velocity is governed
by behavior patterns of economic agents and open a way to
constructing micro-foundation of it. It is found that time pattern
of income and expense for a representative individual can be
obtained from a simple version of life-cycle model, and average
holding time of money resulted from the individual's optimal
choice depends on the expected length of relevant planning
periods.
\end{abstract}

\begin{keyword}
velocity of money \sep holding time \sep life-cycle hypothesis
\sep micro-foundation

\PACS 89.65.Gh \sep 87.23.Ge \sep 05.90.+m \sep 02.50.-r

\end{keyword}

\end{frontmatter}
\section{Introduction}

The topic of money dynamics recently attracted considerable
research interest within statistical physics
\cite{basic,barti,barti2,redner,borti,hayes}. Motivated by the
initial work of Vilfredo Pareto in 1897  on distribution of income
among people in some countries\cite{pareto} and some recent
empirical observations of econophysicists
\cite{emp1,emp2,emp3,emp4}, several money transfer models have
been developed for the equilibrium money distribution based on the
analogy between market economics and kinetic theory of gases
\cite{basic,barti,barti2,redner,emp3}. Adrian Dr\"{a}gulescu and
Victor Yakovenko proposed a model of agents exchanging randomly in
a closed economy and demonstrated that the equilibrium probability
distribution of money follows the exponential Boltzmann-Gibbs law
due to time-reversal symmetry and conservation of money
\cite{basic}. The research group of Bikas K. Chakrabarti
investigated the statistical mechanics of money similarly,
considering effects of the saving behavior of the agents
\cite{barti,barti2}. Based on these studies of the subject, Wang
\textit{et.al.} introduced a new aspect of the problem: the
probability distribution of money holding times \cite{wang,ding}.
These works expand the scope of such investigations from focusing
on the static distribution of money to discussing the circulation
of money.

The velocity of money circulation is a central matter in monetary
theory which has attracted much attention for hundreds of
years\cite{barnett}. Although exploration of the velocity can be
traced backward to the earlier works in 1660s\cite{humphrey}, most
of current investigations of velocity is commonly attributed to
Irving Fisher who presented an influential exchange equation $MV =
PY$, where M is the amount of money in circulation, V is the
velocity of circulation of money, P is the average price level and
Y is the level of real income \cite{fisher}. From this equation,
the velocity of money can be computed as the ratio of transaction
volume or aggregate income to money stock. Basing on this
equation, many theoretical and empirical research works on the
velocity have been carried out to examine its
determinants\cite{laidler,bridel,friedman1,friedman2}. However,
the Fisher Equation just proposed a definition and a measurement
of velocity in any way, it could not uncover the intrinsic
properties of the velocity. It resembles Ohm's law, though with
the help of the equation $V = RI$, resistance $R$ can be
calculated by dividing supplied voltage $V$ by consequent current
$I$, it is determined by its internal characteristics, rather than
those two variables. Just as Rothbard\cite{rothbard} reminds us,
``it is absurd to dignify any quantity with a place in an equation
unless it can be defined independently of the other terms in the
equation.''  Thus, by analogy with resistance, the velocity must
have a ``life of its own'' characterized by the money holders.

The purpose of this paper is to analyze the circulation velocity
of money on the base of micro analysis. In our recent work, the
velocity was expressed as a statistical function of holding time
of money\cite{wang}. This relation provides a novel channel
through which the velocity can be determined by individual
goal-directed behavior. With the help of the life-cycle
hypothesis, an alternative approach without taking any aggregate
macroeconomic variables into account is presented for examining
the monetary velocity.

\section{Holding Time versus Velocity of Money}
To use money, people must hold money. However, what they are
holding now must be spent on something on certain moment later, so
a given amount of money can be used again and again to finance
people's purchases of goods and services. In other words, the
money one person spends for goods and services at any given moment
can be used later by the recipient of that money to purchase yet
other goods and services. As a result, there must be an interval
of time between the receipt of income and its disbursement over
which money is held.

This interval was ever called the ``average period of idleness''
or ``interval of rest'' of money by Wicksell\cite{wicksell}. It is
obvious that the interval is not a character of money itself, but
a character of its holder's behavior in utilizing the money. As
Mises\cite{mises} has made clear, ``money can be in the process of
transportation, it can travel in trains, ships, or planes from one
place to another. But it is in this case, too, always subject to
somebody's control, is somebody's property.'' So we called it
holding time of money, placing emphasis on the prefix
\textit{holding}.

The length of holding time is determined by the holder's motives
to use money which are usually sorted into three types within the
theory of money demand: the transactions-motive, the
precautionary-motive and the speculative-motive\cite{15}. At any
given moment, money in an economy may have different holding
times, due to either different holders or various motives of
holding money.

It is reasonable to assume that money spreads over the domain of
holding times ($0,\infty $). Under continuous assumption, there
must be a portion of money $P(\tau )\,d\tau $ whose holding time
is between $\tau $ and $\tau +\,d\tau .$ $P(\tau )$ is defined as
the probability density of money which has the following property

\begin{equation}
\int\nolimits_{0}^{\infty }P(\tau )\,d\tau =1.  \label{normal}
\end{equation}

We know that the distribution of money over the holding time in an
economy evolves with time. That is to say, the holding time of the
same unit of money may be altered at different moment due to the
alteration of holder. Under certain conditions we can get an
equilibrium state where the distribution of money over holding
time is stationary. In this state, any single money's holding time
may strongly fluctuate over time, but the overall probability
distribution of money does not change.

Given a stationary distribution of money over the holding time, we
then have two variables which are related to each other. The first
one is average holding time of money $\bar{\tau}$ which is defined
as the expectation of the holding time of money, i.e.,
\begin{equation}
\bar{\tau}=\int\nolimits_{0}^{\infty }P(\tau )\,\tau \,d\tau .
\label{averagetau}
\end{equation}%
The second one is the circulation velocity of money which can be
expressed by the mathematical expectation of $1/\tau $ as
follows\cite{ding}

\begin{equation}
V=\int\nolimits_{0}^{\infty }P(\tau )\,\frac{1}{\tau }\,d\tau .
\label{velocity}
\end{equation}%
It follows immediately that the velocity is inversely related to
the average holding time.

Since money is identical for participating trade in an economy, it
is always spent randomly in exchange process. Thus the exchange
process for money is of Poisson type, and the probability density
function of money takes the following form\cite{wang}

\begin{equation}
P(\tau )=\lambda e^{-\lambda \tau },  \label{Gama}
\end{equation}%
where $\lambda $ corresponds to the intensity of the Poisson
process. With a few manipulations we get
\begin{equation}
V=\frac{1}{\bar{\tau}}.  \label{relation}
\end{equation}%
This simple and explicit expression gives an intuitive view of the
inverse relation between the velocity and the average holding
time: the longer average holding time of money, the smaller the
velocity of money, and vice versa.

At first glance, this relationship does not bring forth anything
new since holding time and velocity of money are just different
sides of the same coin. However, the relationship implies that all
the impacts of factors associated with the velocity must takes
their effects through influencing people's choices of holding
time. Thus, the velocity of money could be investigated by
examining how economic agents make decisions on the holding time
of money. As we will illustrate in the later sections, this
viewpoint of monetary circulation can open a new venue to
identifying the determinants of the velocity.

An economic system consists of various agents who may utilize
money, such as consumers, firms, and investors etc. Each kind of
agents has its own character when they make economic decisions
involved with money. To calculate the average holding time of
money in an economy, we should first consider separately various
behavior patterns of these agents according to their respective
motives of holding money, and then put them together. The main
purpose of this paper is to show how the holding time can link the
economic decisions and the velocity of money. Thus We choose only
one kind of agents, consumers, as a representative to examine how
their spending decisions could be reflected by the velocity. The
available model for analyzing such kind of agents is the
life-cycle hypothesis which is well-known in economics.

\section{The Life-Cycle Model}

The popular life-cycle hypothesis was produced by Franco
Modigliani, Richard Brumberg and Albert Ando which has been so far
used as a theoretical basis for many empirical
investigations\cite{modigliani}. The primary theme of this
hypothesis is to examine why people save and why the amount they
do. The model considers a representative individual who expects to
live $T$ years more, whose lifetime utility is
\begin{equation}
U=\int\nolimits_{0}^{T}u(C(t))\,dt,  \label{utility}
\end{equation}
where\ $u(\cdot )$ is the instantaneous utility function with the
following properties

$u^{\prime }(\cdot )>0,u^{\prime \prime }(\cdot )<0;$

and $C(t)$ is consumption in time $t$.\ The individual has initial
wealth of $W_{0}$ and expects to earn labor income $Y(t)$ in the
working periods of his or her life. The individual can save or
borrow at an exogenous interest rate, subject only to the
constraint that any outstanding debt must be repaid at the end of
his or her life. For simplicity, the interest rate is set to zero.
Thus the individual's budget constraint is

\begin{equation}
\int\nolimits_{0}^{T}C(t)\,dt\leq
W_{0}+\int\nolimits_{0}^{T}Y(t)\,dt. \label{constraint}
\end{equation}

The individual is maximizing the lifetime utility Eq.
(\ref{utility}) subject to the budget condition Eq.
(\ref{constraint}). We can solve this constrained optimization
problem by using the Lagrangian method. The corresponding
Lagrangian is given by

\[
L=\int\nolimits_{0}^{T}u(C(t))\,dt+\theta \lbrack
W_{0}+\int\nolimits_{0}^{T}Y(t)\,dt-\int\nolimits_{0}^{T}C(t)\,dt],
\]

where $\theta $ is the Lagrange multiplier. The first-order condition for $%
C(t)$ is

\begin{equation}
u^{\prime }(C(t))=\theta .  \label{firstcond}
\end{equation}

Since (\ref{firstcond}) holds in every moment, the marginal
utility of consumption is constant. And since the level of
consumption uniquely determines its marginal utility, this means
that consumption must be constant. Thus we have $C(t)=C$ for all
$t$. Substituting this fact into the budget constraint yields

\begin{equation}
C=\frac{1}{T}[W_{0}+\int\nolimits_{0}^{T}Y(t)\,dt].
\label{consumption}
\end{equation}

The results indicate that people want to smooth their consumption
even though their income may fluctuate in their life time.

It is worth to note that the life-cycle hypothesis is a purely
microeconomic theory.  Along with ``permanent income'' hypothesis
which was formulated by Milton Friedman\cite{friedman3}, it builds
microeconomic footings for the consumption function in which
behavior is goal-directed. Thus this hypothesis can be regard as
the micro-foundation of total consumption which is a part of
aggregate demand. In this model, we can see that no aggregate
variable appears, so according to the exchange equation proposed
by Fisher it seems that the life-cycle hypothesis could not
provide anything useful for analyzing the velocity of money.
However, as we will illustrate below, with the help of the concept
of holding time of money, the micro-foundation of velocity can
also be constructed upon this hypothesis.

\section{Results and Discussion}

For simplicity of exposition, we adopt a very simple version of
life-cycle model where an individual will work for only $T_{0}$
periods in his life, within which his labor income $Y(t)=Y$, and
has no income in the left retiring periods. Assuming the
individual begins with assets of $W_{0}=0$ and dies with assets of
$W_{T}=0$, he will then have a amount of $W=YT_{0}$ that can be
spent in his entire life. From Equation (\ref{consumption}), we
can see that he will be able to spend $C=YT_{0}/T$ each time unit.

In this case, money is the sole form of the individual's assets,
and there is a complete receipt-payment process of money in the
individual's entire life, in which the income $Y(t)$ is the inflow
of money and consumption expending $C(t)$ is the outflow. Thus the
average holding time of money corresponds to the delay time in
this process, which can be computed by the following form

\begin{equation}
\bar{\tau}=\frac{\int\nolimits_{0}^{T}[C(t)-Y(t)]t\,dt}{\int%
\nolimits_{0}^{T}Y(t)\,dt}.  \label{delaytime}
\end{equation}

In the model above the average holding time in the period of
lifetime can thus be obtained as follows

\begin{equation}
\bar{\tau}=\frac{1}{2}(1-\frac{T_{0}}{T})T.  \label{avdelay}
\end{equation}

According to Equation (\ref{relation}) we then get

\begin{equation}
V=\frac{2}{T-T_{0}}.  \label{valueofv}
\end{equation}

From this result we can see that the velocity of money depends
reversely on the difference between the expected length of life
and that of working periods. The larger the difference is, the
smaller the velocity, and vice versa.

Now consider a special case directly follows from Equations
(\ref{avdelay}) and (\ref{valueofv}). When $T_{0}=T$ which means
that there is no retirement for the individual, so that we have
$\bar{\tau}=0$ and $V\rightarrow \infty $. This implies that as
soon as people receive the money they would pay them out
immediately. This result is obviously unreasonable, since it
originates from a strict assumption that income is injected into
people's account consecutively as a flow. However, people get
their salary or income discretely in reality. So we can assume
that total wealth the individual gets in his life is as the same
as above, i.e., $W=YT_{0}$, but the income is paid at equal time
intervals to the individual. Assuming that there is $n$ times
income paying in the working periods, then at each time the amount
he gets is $YT_{0}/n$, and the average interval for paying is
$T_{s}=T_{0}/n$. After these assumptions being considered, the
labor income takes the following form

\begin{equation}
Y(t)=\frac{YT_{0}}{n}\delta (t-kT_{s}),  \label{disincome}
\end{equation}%
where $\delta(\cdot)$ is delta function and $k=0,1,\cdots ,n-1.$

In this case, according to Equation (\ref{consumption}) the
individual's consumption over the lifetime is still smooth.
Substituting these facts into Equation (\ref{delaytime}), we
obtain

\begin{equation}
\bar{\tau}=\frac{1}{2}(1-\frac{T_{0}}{T})T+\frac{1}{2}T_{s}.
\label{avplantime}
\end{equation}

From this result we can see even when $T_{0}=T$ the average
holding time becomes $\bar{\tau}=\frac{1}{2}T_{s}$ rather than
$\bar{\tau}=0$, in accordance with the reality.

The results we have obtained in this model indicate that the
velocity depends on neither amplitude of income nor other
characteristics of the individual, but only the time pattern of
income and expending. Thus we can conclude that the velocity is an
independent entity and has a ``life of its own.'' Although the
velocity is always value of aggregate transactions divided into
money stock, it is the velocity which determines the aggregate
transactions, rather than the reverse.

Even though this conclusion is restricted by the specific
assumptions of the life-cycle model, it still holds by altering
and/or adding some other variables when the assumption is relaxed
into a more general case. What this simple version of life-cycle
model conveys us is how people distribute their income to their
spending plan. This plan can last for a certain period such as
whole life in this model, it can also last for a shorter or a
longer one. Meanwhile, the income can be allocated to various
spending plans which may have different planning periods. Whether
the time period of plan is long or short, so long as the time
pattern of income is given, the time pattern of consumption
expenditure can be obtained by maximizing the representative
individual's goal. Then the average holding time is given by
Equation (\ref {delaytime}). As a result, the micro-foundation for
velocity of money has been constructed.

Strictly speaking, the velocity we obtained above corresponds to
consumption velocity rather than the ratio of GDP (gross domestic
product) to the money stock. In order to cover the whole velocity,
other main components of GDP, such as investment and government
spending, should also be included. As mentioned above, the average
holding time of money in the whole economy is determined not only
by consumers' but also the other economic agents' behaviors.
Whatever the economic agents are, according to the method we
provide above, the corresponding average holding time can be
deduced from their own optimal choice. Therefore, the theories of
consumption and investment in macroeconomics provide
micro-foundation not only for aggregate demand but also for the
circulation velocity of money.

\section{Summary}
We have proposed an alternative approach for understanding the
velocity of money circulation on the base of micro analysis. A
concept of holding time of money has been introduced to reflect
money holders' behavior. It is found that the velocity of money
can be expressed by a mathematical expectation value of the
inverse of holding time of money. The velocity and the average
holding time are reversely related to each other and the main
determinant of them is the agents' behaviors. This relationship
suggests that the micro-foundation of the velocity can be
constructed by examining a representative individual behavior.
Employing the life-cycle hypothesis, We demonstrate that the
average holding time can be obtained by maximizing the
representative individual's lifetime utility. The velocity is
found to be determined by the time pattern of income and dispense
which results from the individual's optimal choice. We believe
that this study promises a fresh insight into the velocity of
money circulation.

\section*{Acknowledgments}
This research was supported by the National Science Foundation of
China under Grant No. 70371072 and 70371073. The authors
benefitted from collaboration with Ning Ding and Li Zhang. We are
also grateful to Zhonghua Cai for helpful discussions, suggestions
and comments.

\end{document}